\documentclass[structabstract]{aa}  
\usepackage{graphicx}
\usepackage{epstopdf}
\usepackage{txfonts}

\begin{document}

\title{Phase sensor for solar adaptive-optics}

   \author{A. Kellerer 
          }

   \institute{
   Big Bear Solar Observatory, \\
   40386 North Shore Lane\\
Big Bear City CA 92314-9672, USA\\
              \email{kellerer@bbso.njit.edu}
             }


  \abstract
 { Wavefront sensing in solar adaptive-optics is currently done with correlating Shack-Hartmann sensors, although the spatial- and temporal-resolutions of the phase measurements are then limited by the extremely fast computing required to correlate the sensor signals at the frequencies of daytime atmospheric-fluctuations.}
{To avoid this limitation, a new wavefront-sensing technique is presented, that makes use of the solar brightness and is applicable to extended sources.} 
{The wavefront is sent through a modified Mach-Zehnder interferometer. A small, central part of the wavefront is used as reference and is made to interfere with the rest of the wavefront. }
{The contrast of two simultaneously measured interference-patterns provides a direct estimate of the wavefront phase, no additional computation being required. 
The proposed optical layout shows precise initial alignment to be the critical point in implementing the new wavefront-sensing scheme.}


   \maketitle

\section{Introduction}

Adaptive-optic corrections for solar observations are currently based on Shack-Hartmann (SH) wavefront sensors, see e.g. Rimmele (\cite{Rimmele}). 
Such sensors divide the wavefront by an array of lenslets, and each wavefront element forms a separate image on the detector. Local wavefront slopes are inferred from the image positions. When the source is a distant star, the images are small discs (either diffraction- or seeing-limited) and the image-centers are computed through a barycenter calculation. 
For solar adaptive-optics, however, the image behind each lenslet is extended and the displacements must, therefore, be computed through a cross-correlation algorithm. 

With solar observations the number of lenslets is not limited by flux considerations, as with the much fainter night-time targets, but by computational constraints. Even with the fastest available computers the calculations need to be optimized to meet the necessary correction-rates of $\sim 2$kHz. Thus, at the {\it New Solar Telescope (NST)\/} in Big Bear, California, the data from the 76 sub-aperture SH-sensor are analyzed with digital signal processors (DSP), programmed in assembly language, and the correlation is reduced to displacements within $\pm 3$ pixels, see e.g. Rimmele et al. (\cite{Rimmele2}), Richards et al. (\cite{Richards2004}, \cite{Richards2008}) and Denker et al. (\cite{Denker}). Suitable algorithms are currently being developed for the planned upgrade to 308 sub-apertures. 
At the German {\it Vacuum Tower Telescope\/} (VTT) in Tenerife, Spain, the use of fast Fourier transforms permits correction frequencies of 2100\,Hz on a total of 36 sub-apertures. For the GREGOR telescope, Tenerife, Spain, the same team develops algorithms to sense the wavefronts at 2500\,Hz on 156 sub-apertures (Berkefeld et al.,\,\cite{Berkefeld}). 

A new wavefront-sensing technique is presented here. In this approach the phase is inferred without the need for a cross-correlation algorithm, and the computation requirements are greatly relaxed. A small, central part of the wavefront is used as reference and is made to interfere with the rest of the wavefront. A phase shift of $\pi/2$ is  introduced in one arm of the interferometer, and the two beams are then recombined via a beam-splitter cube. The wavefronts interfere with a phase difference from their common mean-value which varies around $\pi/2$ in one output, and $-\pi/2$ in the other output. The normalized intensity-difference between the two outputs  is a direct measure of the wavefront phase. No additional computation is required. 

The method makes use of the Sun's brightness and is applicable to extended sources. 
Since all phases are measured relative to the same, central phase-value, the error propagation is independent of the number of sub-apertures.  

\section{Formalism}

Let $\vec{E}$ be the electric field associated with the wavefront:
\begin{eqnarray}
\vec{E}(x,y)=\vec{E}_0\cdot P(x,y) \cdot \exp(i\,\phi(x,y))\cdot \exp(i\,(kz-\omega t))
\end{eqnarray}
The wave propagates along the $z$-axis in the orthogonal coordinate system $(x,y,z)$. $k=2\pi/\lambda$ is the wave number, $\lambda$ the wavelength. $\omega=2\pi c/\lambda$ denotes the angular frequency, $c$ being the speed of light. $\phi$ is the wavefront phase and $P$ the pupil transmission function. We consider a circular pupil of diameter $D$:
\begin{eqnarray}
P(x,y) &=& 1 \hspace{0.4cm} {\rm if\/}\hspace{0.2cm} x^2 + y^2 \leq \frac{D^2}{4} \nonumber \\	
&=& 0 \hspace{0.4cm} {\rm otherwise\/}
\end{eqnarray}

In the proposed method, the wavefront is sent through a modified Mach-Zehnder  interferometer: 
the central part of the wavefront with diameter $d=D/N$ is propagated along path (A), the complement is propagated along path (B). 

\begin{itemize}
\item[--] In path (A), an achromatic afocal system enlarges the central wavefront from diameter $d$ to $D=N\,d$. 
Behind the afocal system the electric field equals:
\begin{eqnarray}\label{eq:EA}
\vec{E}_A(x,y)&=& \frac{\vec{E}_0}{N^2}\cdot P(x,y) \nonumber \\
&&\cdot \exp \left(i\,\phi(\frac{x}{N},\frac{y}{N})\right)\cdot \exp\left(i\,(kz-\omega t)\right)
\end{eqnarray}

\item[--] Path (B) introduces an achromatic $\pi/2$ phase-shift relative to path (A), and attenuates the amplitude of the electric field by a factor $N^2$ in order to make it comparable to path (A):  
\begin{eqnarray}\label{eq:EB}
\vec{E}_B(x,y)&=& \frac{\vec{E}_0}{N^2}\cdot P'(x,y) \nonumber \\
&&\cdot \exp\left(i\,(\phi(x,y)+\pi/2)\right)\cdot \exp\left(i\,(kz-\omega t)\right)
\end{eqnarray}
$P'$ is the transmission function in path (B): $P'(x,y) = P(x,y) - P(x\, N,y\, N)$. 
\end{itemize}

The two beams are recombined via a beam splitter, and the electric fields in the two outputs are then expressed by:
\begin{eqnarray}
\vec{E}_1(x,y)&=& \frac{1}{\sqrt 2} ( \vec{E}_A(x,y) + \vec{E}_B(x,y))\nonumber \\
\vec{E}_2(x,y)&=& \frac{1}{\sqrt 2} ( \vec{E}_A(x,y) + \exp(i\pi)\cdot \vec{E}_B(x,y))
\end{eqnarray}\label{eq:E12}
The additional phase-delay, $\pi$, for the field propagating from path (B) into output (2) is introduced upon reflection on the front of the beamsplitter mirror: the medium behind the mirror is glass, and has a higher refractive index than the air the field is traveling in.
The field propagating from path (A) into output (1) reflects on the mirror-back, inside the glass of the beamsplitter and no additional phase shift is thus introduced.

The resulting intensities are: 
\begin{eqnarray}
I_1(x,y)&=& \frac{E_0^2}{N^4} \cdot P'(x,y) \left[1+\cos\left(\phi(x,y)+\pi/2-\phi(\frac{x}{N},\frac{y}{N})\right) \right] \nonumber  \\
I_2(x,y)&=& \frac{E_0^2}{N^4} \cdot P'(x,y) \left[1+\cos\left(\phi(x,y)-\pi/2-\phi(\frac{x}{N},\frac{y}{N})\right) \right] 
\end{eqnarray}\label{I12}
These expressions do not apply inside the central disc of diameter $d$, where the field $\vec E_B$ has zero amplitude and where both intensities equal: $I_1(x,y)=I_2(x,y)= E_0^2/N^4$. In the following, only those points $(x,y)$ are considered where $P'(x,y)=1$. 

The central wavefront element -- sent through path (A) -- is used as reference. We therefore make the following approximation: 
\begin{eqnarray}
\phi(\frac{x}{N},\frac{y}{N}) \sim \phi(0,0) = 0 , \hspace{0.9cm}{\rm if\/}\hspace{0.2cm} x^2 + y^2 \leq \frac{D^2}{4} 
\end{eqnarray}\label{eq:I12}

So that, 
\begin{eqnarray}
I_1(x,y)&\sim&  \frac{E_0^2}{N^4} \cdot P'(x,y) \left[1- \sin\left(\phi(x,y)\right)\right] \nonumber  \\
I_2(x,y)&\sim&  \frac{E_0^2}{N^4} \cdot P'(x,y) \left[1+\sin\left(\phi(x,y)\right)\right]
\end{eqnarray}

Under the condition of small phase distortions $\sin(\phi)\sim\phi$, and the intensity-contrast equals:
\begin{eqnarray}\label{eq:C}
\frac{I_1(x,y)-I_2(x,y)}{I_1(x,y)+I_2(x,y)}  \sim \phi(x,y)
\end{eqnarray}

The phase at position $(x,y)$ is inferred from only two intensity measurements, $I_1(x,y)$ and $I_2(x,y)$. The error propagation from measurement to phase-estimate is thus independent of the number of measurement points (sub-apertures), while it increases logarithmically with the number of sub-apertures when SH-sensors are employed (Hudgin\,\cite{Hudgin}, Fried\,\cite{Fried}, Kellerer \& Kellerer\,\cite{AKAK}).

Turbulence-induced path-length distortions, $\delta$, are achromatic, hence phase distortions are inversely proportional to the wavelength: $\phi(x,y,\lambda)=2\pi \,\delta(x,y)/ \lambda$. If the spectral range of the phase sensor is $[\lambda_0,\lambda_0+\Delta\lambda]$, Eq.\,\ref{eq:C} becomes: 
\begin{eqnarray}
\frac{I_1(x,y)-I_2(x,y)}{I_1(x,y)+I_2(x,y)} & \sim & \frac{2\pi}{ \Delta\lambda} \cdot \ln(1+\Delta\lambda/\lambda_0) \cdot \delta(x,y)\nonumber \\
&&
= \phi(x,y)\sum_{i=0}^{+\infty}\frac{(-1)^i}{i+1} \left(\frac{\Delta\lambda}{\lambda_0}\right)^i
\end{eqnarray}
Wavefront sensing in solar adaptive-optics is typically done inside a 50\,nm wide spectral-band centered around 550\,nm, in this case $\sum_{i=0}^{+\infty}(-1)^i(\Delta\lambda/\lambda_0)^i /(i+1) \sim 1$. This band is assumed for our sensor.

\section{Implementation of the phase sensor}

\subsection{Optical Layout}\label{layout}

\begin{figure*}
\centering
\includegraphics[width=.7\textwidth]{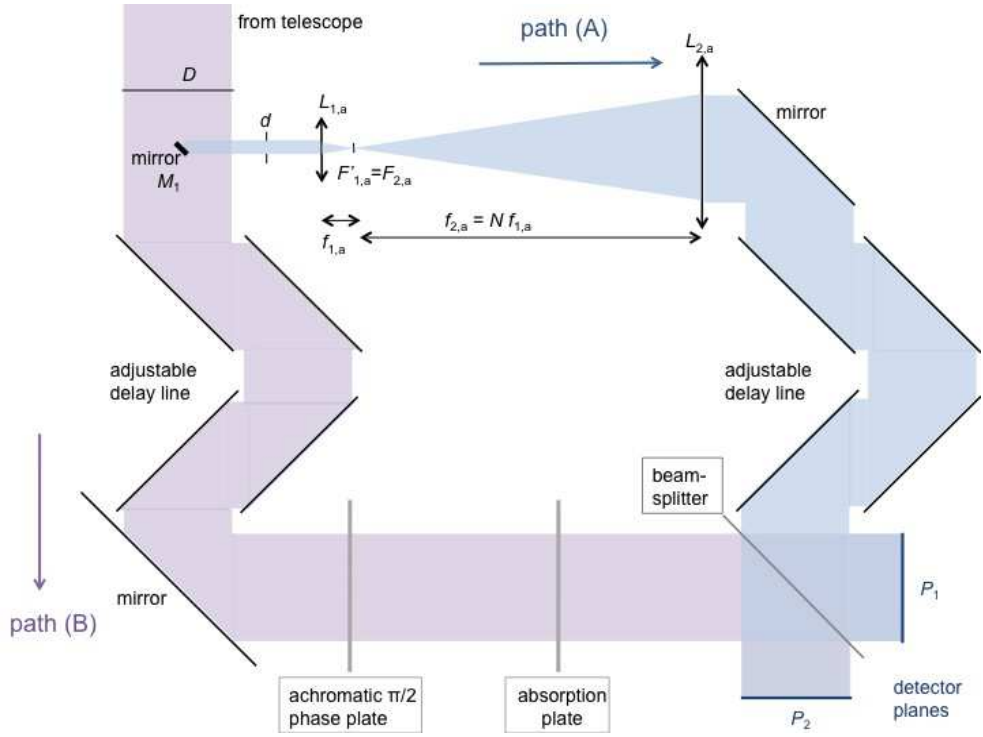}
\caption{Optical layout of the phase sensor. Path (A) propagates the central part of the wavefront. The achromatic afocal system $(L_{1,a}, L_{2,a})$ enlarges the wavefront diameter from $d$ to $D=N\,d$. The two delay-lines compensate for the path-length increment due to the afocal system in path (A). }
\label{fig:1a}
\end{figure*}

Fig.\,\ref{fig:1a} sketches the optical layout as it would be used in the case of a point source.  
Along path (A) the central part of the wavefront is enlarged from its initial diameter $d$ to the pupil diameter $D=N\,d$.
Along path (B) an achromatic phase-retarder introduces a $\pi/2$ phase-shift relative to path (A). An absorption plate attenuates the field-amplitude by a factor $N^2$ in order to approximately equalize the amplitudes in the two paths. 
Two delay lines compensate for the increment in path-length introduced by the achromatic afocal system in path (A). 

The beams are recombined via a beam-splitter. 
For a flat incoming wavefront, the phase difference upon recombination is $\pi/2$ in output (1) and $-\pi/2$ in output (2). 
For a distorted wavefront -- and if the phase variance of the reference wavefront (path A) is negligible -- the phase difference upon recombination equals $(\phi(x,y)+\pi/2)$ in output (1) and $(\phi(x,y)-\pi/2)$ in output (2).

The mean-phase difference introduced by a path-length difference, d$L=L_B-L_A$, between paths A and B equals:  d$\phi=<\phi_B>-<\phi_A>=2\pi \cdot {\rm d}L/\lambda$. A path-length difference controlled within $\sim\lambda/20$ ensures that the instrumentally induced phase differences are well below $\pi/2$, and that the phase differences vary around $\pi/2$ and $-\pi/2$ in the two outputs. 

The method is not actually intended for application to point sources because, on the one hand, the intensity attenuation factor $1/N^2$  would then not be acceptable and, on the other hand,  because any remaining phase fluctuations within the central wavefront segment may, with point sources,  make the broadened beam in path (A) insufficiently uniform. As will be seen, both limitations are less critical in solar observations.

\subsection{Adjustment for an extended object}

\begin{figure*}
\centering
\includegraphics[width=.7\textwidth]{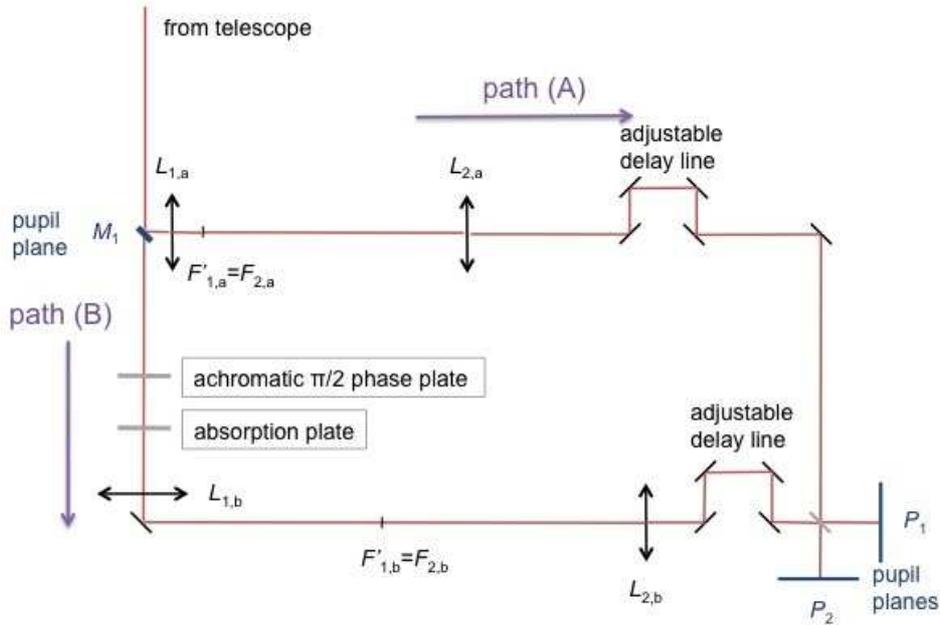}
\caption{Adjustment of the optical layout (Fig.\,\ref{fig:1a}) for an extended source such as the Sun. Both afocal systems, $(L_{1,a}, L_{2,a})$ and $(L_{1,b}, L_{2,b})$, re-image the pupil plane, $M_1$, onto the detector planes, $P_1$ and $P_2$. }
\label{fig:1b}
\end{figure*}

To allow for the image extension in solar observations, the optical layout is adjusted so that the interference patterns from sources in different directions will be properly superimposed (see Fig.\,\ref{fig:1b}):
\begin{itemize}
\item[--]  Mirror $M_1$, which sends the central part of the wavefront into path (A), is placed in a pupil plane, where the wavefronts from different angular-directions overlap. This ensures that the central parts of all wavefronts are picked up by $M_1$. 
\item[--]  The afocal system $(L_{1,a}, L_{2,a})$ is adjusted to re-image the pupil plane ($M_1$) onto the two detectors ($P_1$ and $P_2$). 
This requires $L_{1,a}$ and $L_{2,a}$ to be convergent lenses, and $M_1$ to be placed between $L_{1,a}$ and the focal plane of $L_{1,a}$. If $L_{1,a}$ were divergent and $L_{2,a}$ convergent, the image of $M_1$ through $(L_{1,a}, L_{2,a})$ would be virtual, and could, thus, not be projected onto the detector planes, $P_1$ and $P_2$.   
\item[--]  Another afocal system $(L_{1,b}, L_{2,b})$ is placed in path (B) to re-image the pupil-plane, $M_1$, onto the two detectors, $P_1$ and $P_2$. 
\end{itemize} 
 
With the adjusted optical layout, the measured contrast equals the intensity weighted phase-average over a solid-angle, $\Omega$: 
\begin{eqnarray}\label{eq:ext}
\frac{I_1(x,y)-I_2(x,y)}{I_1(x,y)+I_2(x,y)}  \sim \frac { \iint_\Omega L(\vec\theta)\, \phi(x,y,\vec\theta) \,{\rm d}^2\theta }{  \iint_\Omega L(\vec\theta)\, {\rm d}^2\theta}
\end{eqnarray}
$L(\vec\theta)$\,[m$^{-2}$\,s$^{-1}$\,sr$^{-1}$] is the solar radiance from the angular direction $\vec\theta$.  

Adaptive optics with a single deformable mirror is in principle limited to acceptance angles within the isoplanatic diameter ($\alpha=10"$ for typical daytime observations). For larger acceptance angles, $\Omega >\pi \alpha^2$ and the correction can merely account for the mean phase shift over  $\Omega$. 
Since this mean value is predominantly determined by the nearby turbulence, the resulting approximate correction is termed ground-layer adaptive optics. 
A second condition for exact adaptive optics is the sufficiently close spacing of  actuators. Their separation should correspond to not more than the Fried distance (roughly 0.05\,m for typical day-time observations), otherwise the correction accounts again merely for a mean phase shift. 

With a point object, the central sub-aperture contains, depending on its size,  a certain level of phase differences. Ideally one would wish to average out these differences in the widened beam (A) to make it as uniform as possible. While this can not be done, a somewhat increased angle of acceptance will have the similar effect of replacing the phase shifts at each point by a local average which makes the central beam more uniform. The proposed method is, thus, inherently suited for extended images.

\section{Sensitivity of the phase sensor}

\subsection{Maximum apparent magnitude}\label{sec:flux}

Assume that sensor areas equal the size of the central sub-aperture. The broadened beam from the central sub-aperture is then required to interfere with the $N^2-1$ other areas. The number, $Z$, of photons per detector read-out must, thus, be sufficiently high:
\begin{eqnarray}
Z = I\cdot S\cdot \Delta t > Z_0\cdot N^2
\end{eqnarray}
$I$ [m$^{-2}\,$s$^{-1}$] is the target's irradiance, $S$ the collecting surface of the central sub-aperture and $\Delta t$ is the exposure time. $Z_0$ denotes the minimum number of photons for the interference in each sub-area. 
For $Z_0=100$ the signal to photon-noise ratio stays below 10\%. 

Initial calculations (see Section\,\ref{sec:num}) suggest that the root-mean-square (rms) phase-deviation, $\sigma_\phi$, over the whole wavefront needs to remain within $\sim1.2\,$rad. The central part of the wavefront can serve as a reference if its rms phase-deviation, $\sigma_0$, is substantially smaller than $\sigma_\phi$, say $\sigma_0<\sigma_\phi/5$. 
Accordingly, the diameter of the region on the telescope aperture that corresponds to the central sub-aperture should ideally not exceed a fraction of the Fried length, $r_0$ (Roddier\,\cite{Roddier81}):
\begin{eqnarray}
d < r_0\cdot \sigma_{0, \rm max}^{6/5} = 0.19\cdot r_0
\end{eqnarray}
With the diameter of the NST telescope, $D=1.6\,$m, and a Fried length $r_0=0.05$\,m, we then get: $d= 0.01$\,m and $N=D/d=168$. Daytime adaptive-optics correction is typically done at a frequency of $f=1/\Delta t =2$\,kHz, so that: 
\begin{eqnarray}
I>8 \cdot 10^{13} \, \rm{m}^{-2}\,\rm{s}^{-1}
\end{eqnarray}
And the target's apparent-magnitude needs to be below:
\begin{eqnarray}
m < -2.5\,\log \frac{I}{I_0} = -10.5
\end{eqnarray}
$I_0=5\cdot 10^9$m$^{-2}\,$s$^{-1}$ equals the irradiance from a star with zero apparent magnitude inside a 50\,nm wide spectral-band centered around 550\,nm (Allen's Astrophysical Quantities,\,\cite{Allens}). 

When the wavefront sensing is done inside the isoplanatic cone (typically $\alpha=10"$), the corresponding apparent magnitude of the Sun equals: $m=-26.7+2.5\log(0.5\cdot 3600/10)=-21.1$. 
The number of photons received in the central sub-aperture is thus by far sufficient.

 \subsection{Maximum amplitude of the phase-distortions}\label{sec:num}
 
\begin{figure*}
\centering
\includegraphics[width=.3\textwidth]{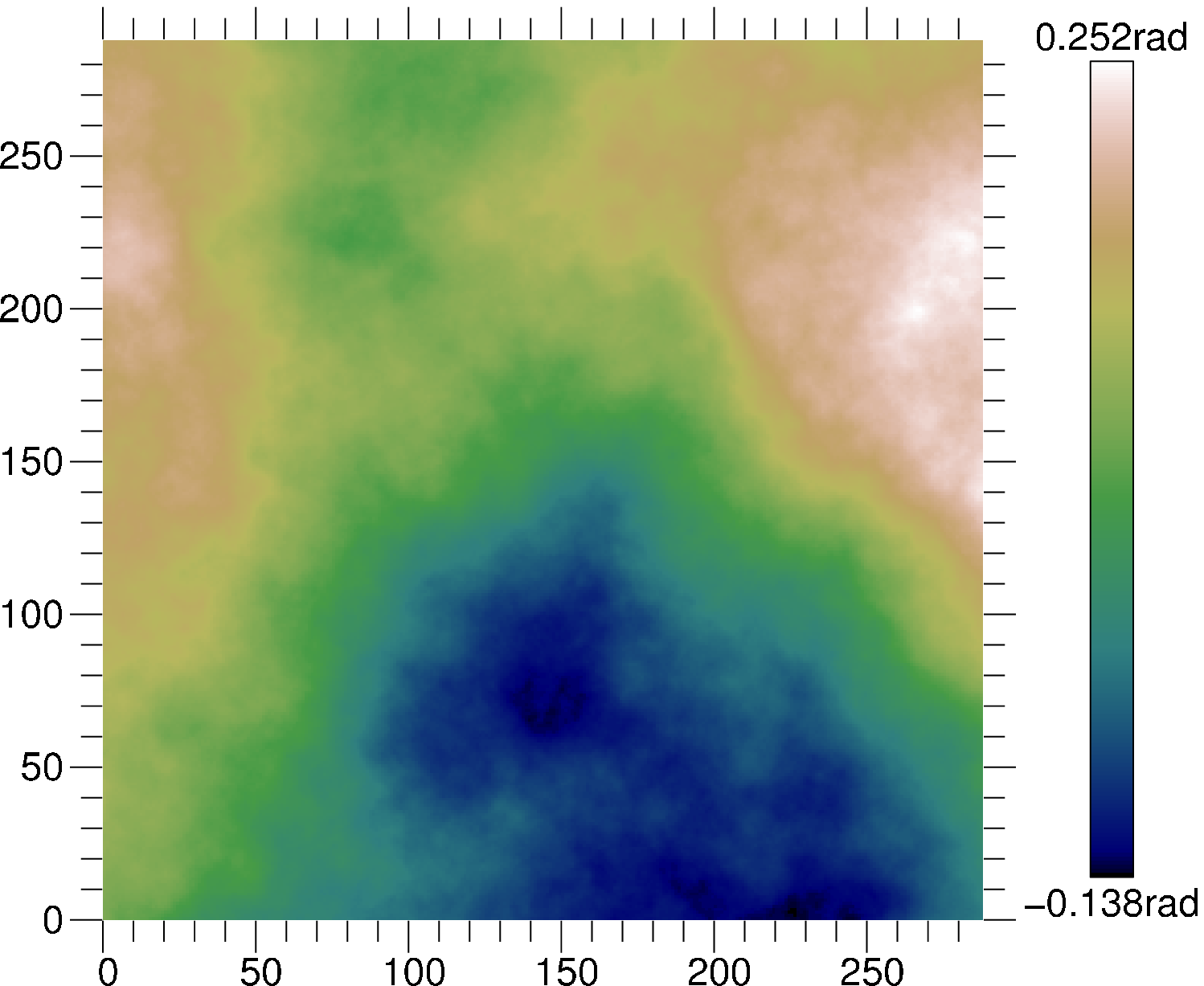}
\includegraphics[width=.3\textwidth]{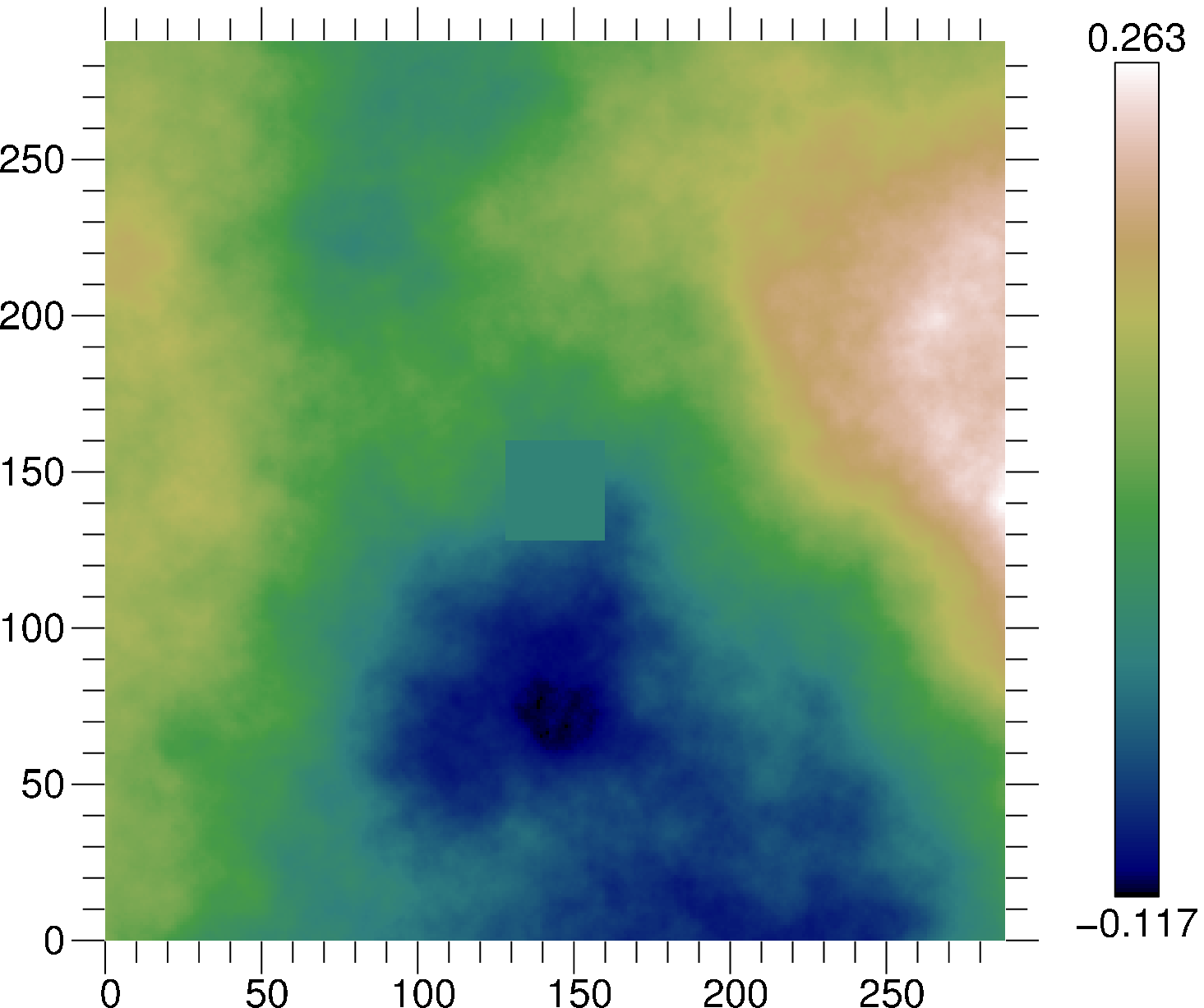}
\caption{Left panel: Matrix of phase values with 0.1\,rad rms deviation. The phase distribution follows the Kolmogorov  statistics. Right panel: Quantity measured by the phase sensor: $(I_1-I_2)/(I_1+I_2)$. }
\label{fig:2}
\end{figure*}

\begin{figure*}
\centering
\includegraphics[width=.3\textwidth]{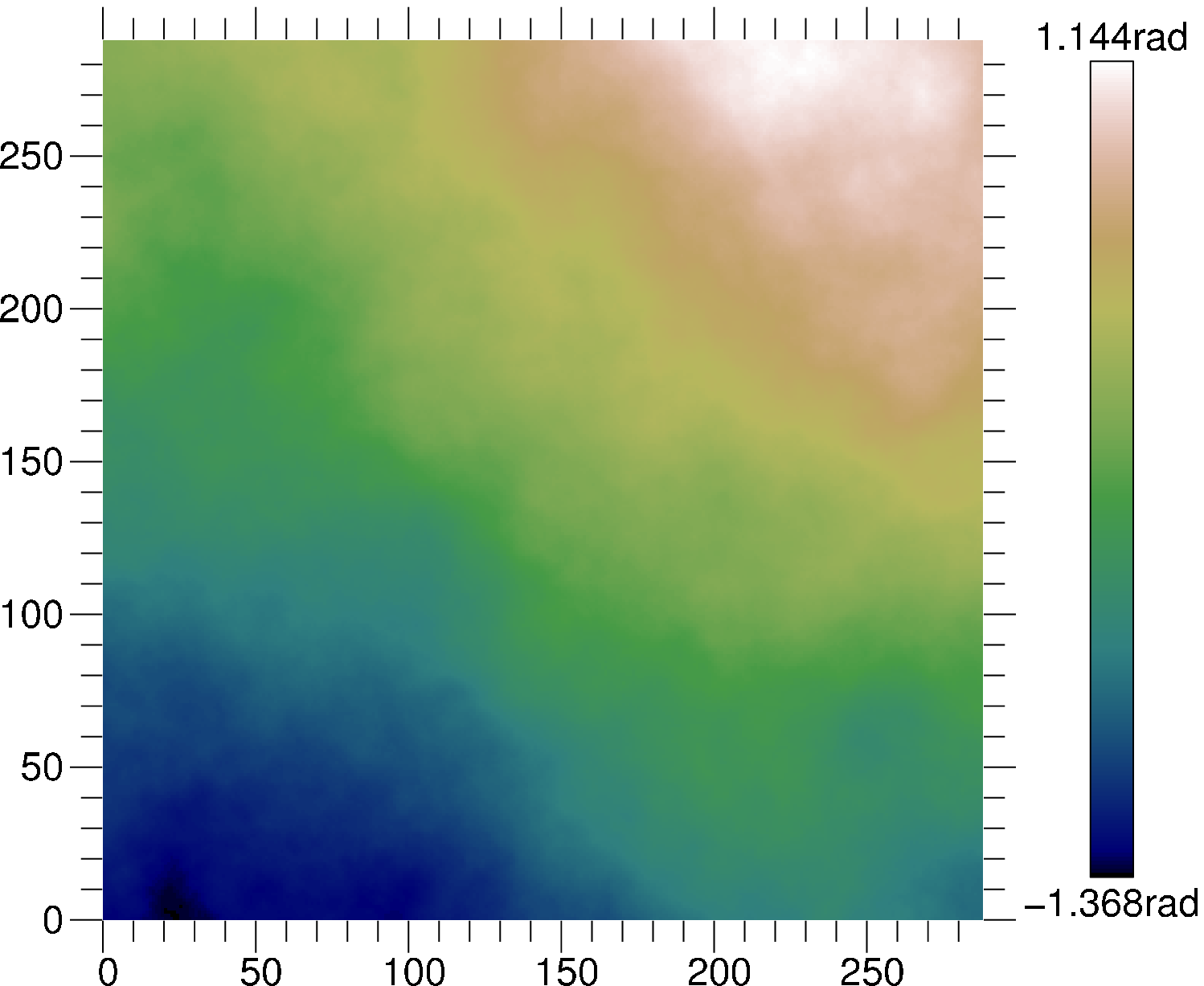}
\includegraphics[width=.3\textwidth]{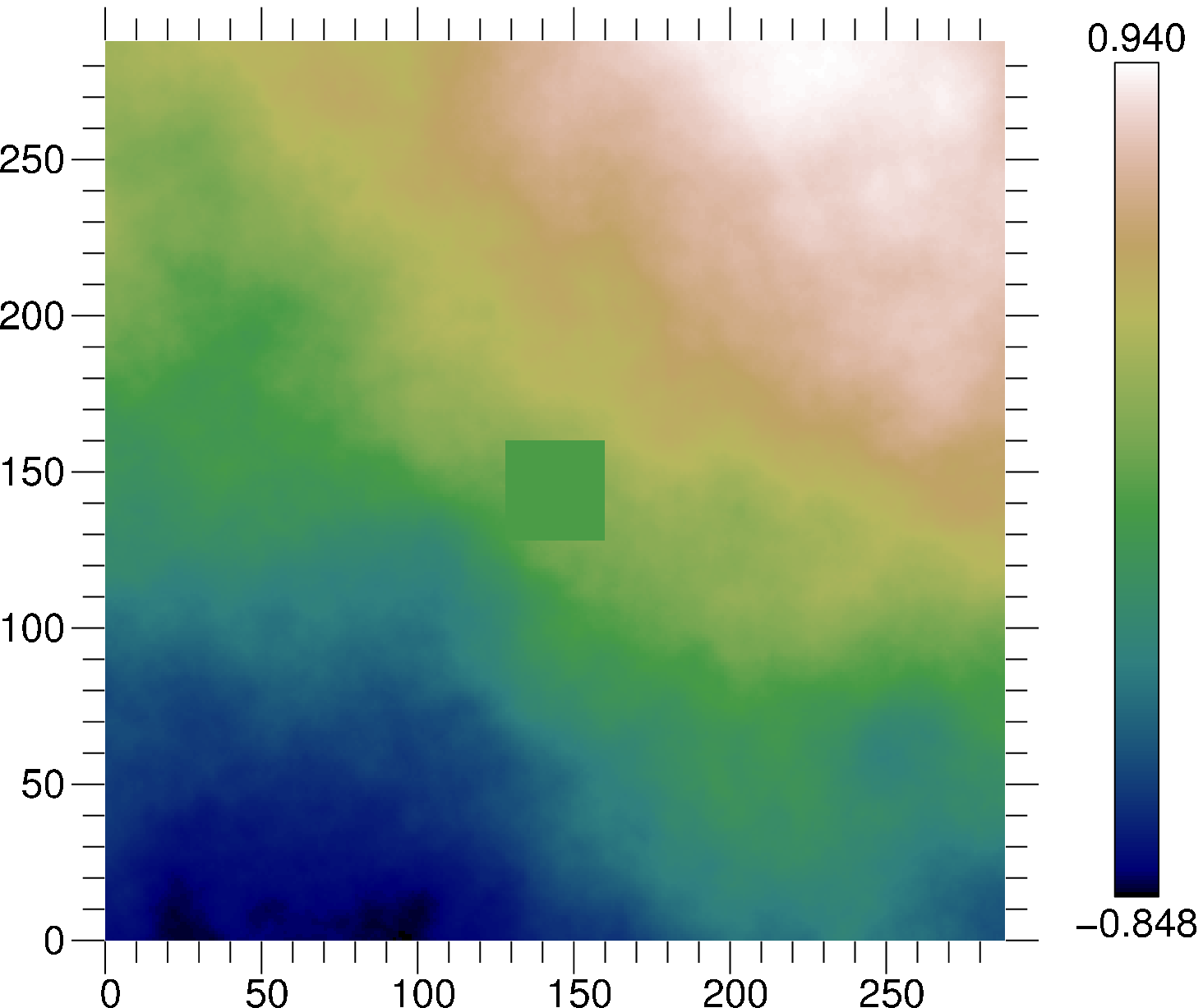}
\caption{Same as on Fig.\,\ref{fig:2}, with a different realization of phase values. The rms phase deviation equals 0.6\,rad. }
\label{fig:3}
\end{figure*}

\begin{figure*}
\centering
\includegraphics[width=.3\textwidth]{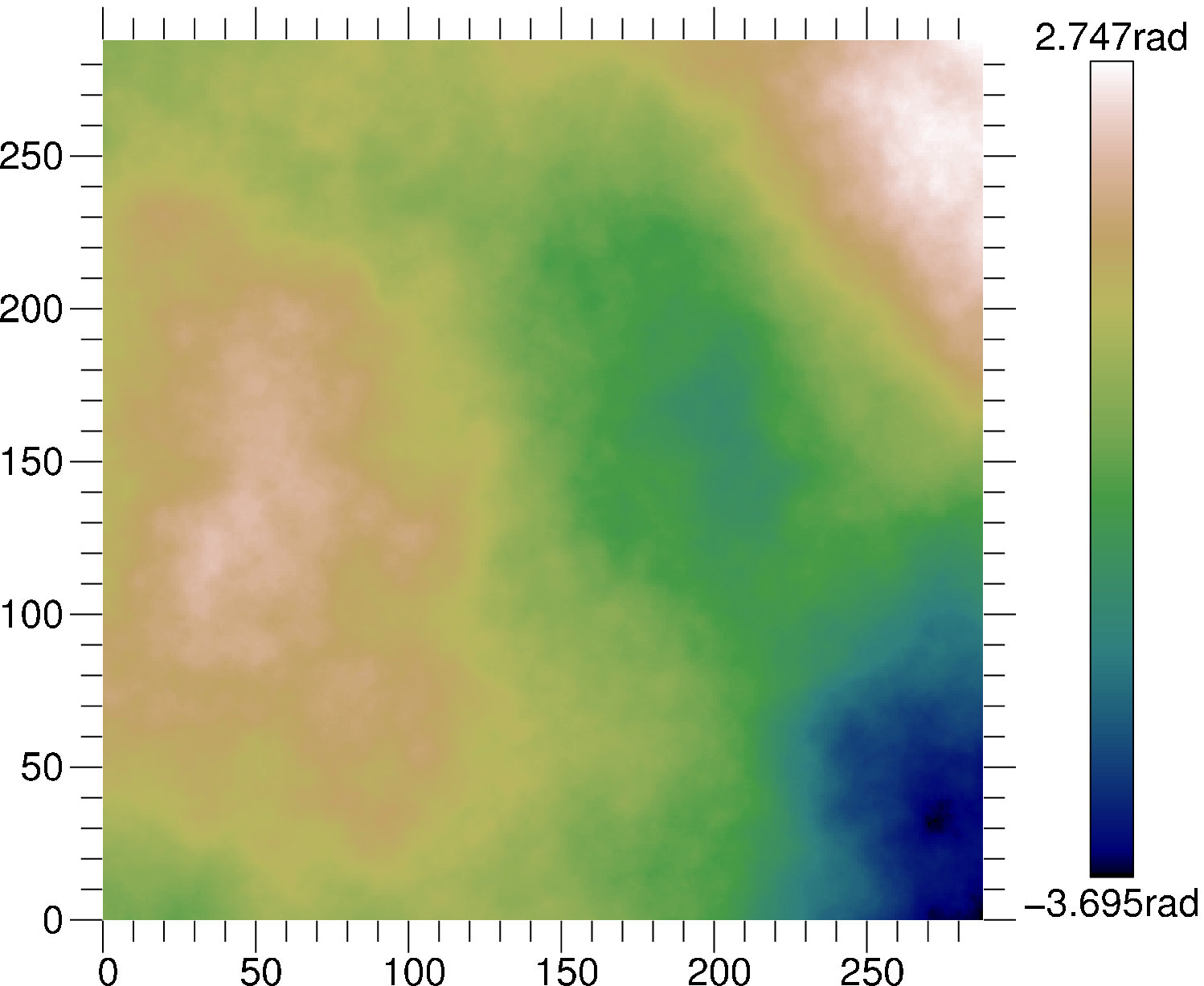}
\includegraphics[width=.3\textwidth]{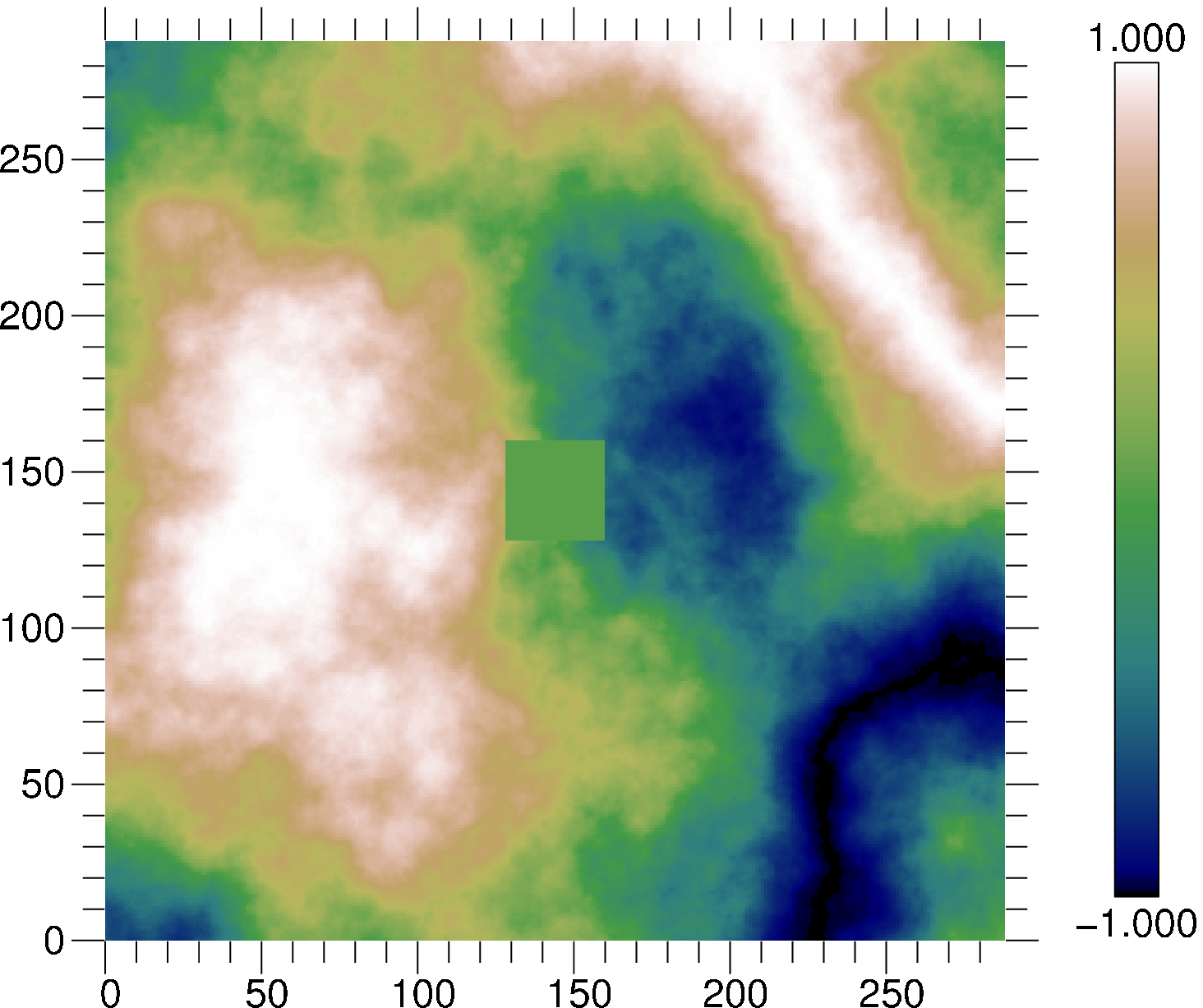}
\caption{Same as on Fig.\,\ref{fig:2}, with a different realization of phase values. The rms phase deviation equals 1.2\,rad.  }
\label{fig:4}
\end{figure*}

In exploratory computations different $288\times 288$ matrices of phase values have been generated with Kolmogorov statistics, and the normalized intensity-differences, $(I_1-I_2)/(I_1+I_2)$, have been calculated in terms of Eqs.\,\ref{eq:EA}--\ref{eq:C}. 
The central $32\times 32$ elements were taken to represent the reference wavefront that is propagated through path (A) of the interferometer. 
The normalized intensity differences are shown in Figs.\,\ref{fig:2}--\ref{fig:4} for different phase screens, with rms phase-deviations between $\sigma_\phi= 0.1$\,rad and 1.2\,rad. 
$\sigma_\phi= 1.2$\,rad appears to be a limit: for larger phase distortions, $\sin(\phi)\sim\phi$ fails to be a valid approximation, and the intensity difference ceases to be proportional to the phase distortion. Instead (see Eq.\,\ref{eq:I12}):  
\begin{eqnarray}
\frac{I_1(x,y)-I_2(x,y)}{I_1(x,y)+I_2(x,y)}  \sim \sin(\phi(x,y))
\end{eqnarray}

The phase values need then to be unwrapped, and the sensing scheme looses its advantage of minimal computational requirements. 
For a $D=1.6$\,m telescope and a Fried length $r_0=0.05$\,m, the turbulence-induced rms phase-distortions equal in fact: $\sigma=(D/r_0)^{5/6}=18$\,rad (Roddier\,\cite{Roddier81}). 
One must, therefore, take into account that the phase differences between beams (A) and (B) are determined only modulo $2\pi$ which requires an initial computational procedure which might start out the adaptive-optics correction-loop on a central subset of the sensor field and progressively extend the set as the correction takes effect, this extension being either algorithmically coded or being controlled by operating a diaphragm in the collimated beam. Once an approximate phase correction is achieved the precise correction can proceed without further computational effort at the desired frequency.

\section{Conclusion}

A scheme is presented to sense wavefront distortions on bright, extended sources. 
The central part of the wavefront is used as reference and is made to interfere with the rest of the wavefront by use of a suitably modified Mach-Zehnder interferometer. The intensity contrast in the two outputs of the interferometer is a direct measure of the wavefront-phases.
Once an approximate phase adjustment is attained the continued exact correction requires no additional calculation, i.e. the  computation requirements are greatly relaxed in comparison to the current use of correlating SH-sensors. 
The most critical step in the implementation of the new wavefront-sensing method is likely to be the initial alignment of the Mach-Zehnder interferometer, since the path lengths in both arms of the interferometers need to be equalized within $\lambda/20$ (see Section\,\ref{layout}). This, as well as the suitable control strategy will need to be explored in an experimental set-up.

The phase differences are deduced from the contrast of two simultaneously measured intensities and, accordingly, the estimates are not affected by scintillation. Furthermore, all phases are measured relative to the same, central phase-value, and the error-propagation factor is therefore independent of the number of sampling elements. This contrasts with the logarithmic increase of  the propagation factor when increasing numbers of SH sub-apertures are used.

The method requires the flux in the central sub-aperture to be sufficiently high, and is thus especially suited for solar adaptive-optics. 

\begin{acknowledgements}
Many thanks go to Nicolas Gorceix for numerous discussions on solar wavefront sensing. 
The  National Science Foundation is also gratefully acknowledged for funding this research through grant NSF-AST-0079482. 
\end{acknowledgements}


\begin{thebibliography}{99}

\bibitem[2000]{Allens}
Allen's Astrophysical Quantities, Ed. A. N. Cox, 4th ed. (2000)

\bibitem[2010]{Berkefeld} T. Berkefeld et al.,  
``Adaptive Optics development at the German solar telescopes'', 
Applied Optics, vol. 49, N. 31, G155--G166 (2007)

\bibitem[2007]{Denker} C. Denker et al.,  
``Adaptive Optics at the Big Bear Solar Observatory: Instrument Description and First Observations'', 
PASP, vol. 119, issue 852, p. 170--182 (2007)

\bibitem[1977]{Fried}  D. Fried,
``Least-square fitting a wave-front distortion estimate to an array of phase-difference measurements'', 
Optical Society of America, Journal, vol. 67, p. 370-375 (1977)


\bibitem[1977] {Hudgin}  R. Hudgin,
``Wave-front compensation error due to finite corrector-element size'',
Optical Society of America, Journal, vol. 67, p. 393-395 (1977)

\bibitem[2011]{AKAK}  A. Kellerer, A. Kellerer,
``Error propagation: a comparison of Shack-Hartmann and curvature sensors'', 
JOSA A, in press (2011)

\bibitem[2004]{Richards2004} K. Richards et al., 
``High speed low latency solar adaptive optics camera'', 
SPIE, vol. 5171, p. 316-325 (2004)

\bibitem[2008]{Richards2008} K. Richards, T. Rimmele,
``Real-time processing for the ATST AO system'', 
Advanced Maui Optical and Space Surveillance Technologies Conference (2008)

\bibitem[2004]{Rimmele} T. Rimmele, 
``Recent Advances in Solar Adaptive Optics'', 
SPIE, vol. 5490, p. 34-46 (2004)

\bibitem[2004]{Rimmele2} T. Rimmele et al., 
``First Results from the NSO/NJIT Solar Adaptive Optics System'', 
SPIE, vol. 5171, p. 179-186 (2004)

\bibitem[1981]{Roddier81} 	
F. Roddier, ``The effects of atmospheric turbulence in optical astronomy'', 		
Progress in optics, vol. 19, Amsterdam, North-Holland Publishing Co., p. 281-376 (1981)




\end{thebibliography}
\end{document}